\documentclass[prb,twocolumn,floatfix,showpacs,superscriptaddress]{revtex4}
\usepackage{amsmath}
\usepackage{graphicx}
\newcommand{\fref}[1]{Fig.~\ref{#1}}
\newcommand{\sref}[1]{Sec.~\ref{#1}}
\newcommand{\aref}[1]{Appendix~\ref{#1}}
\newcommand{\sigmaV}{{{\mbox{\boldmath$\sigma$}}}}

\font\bbfnt=msam10
\def\gsim{\,\hbox{\bbfnt\char'046}\,}
\def\lsim{\,\hbox{\bbfnt\char'056}\,}

\begin{document}

\title{Effect of random interactions in spin baths on decoherence}
\author{S. Camalet}
\author{R. Chitra}
\affiliation{Laboratoire de Physique Th\'eorique de la Mati\`ere Condens\'ee, UMR 7600, Universit\'e Pierre et Marie Curie, Jussieu, Paris-75005, France}

\date{today}
\begin{abstract}
We study the decoherence of a central spin $1/2$ induced by a spin bath with intrabath interactions. 
Since we are interested in the cumulative effect of interaction and disorder, we study 
baths comprising Ising spins with random ferro- and antiferromagnetic interactions between the spins. 
Using the resolvent operator method which goes beyond the standard Born-Markov master equation approach, 
we show that, in the weak coupling regime, the decoherence of the central spin at all times 
is entirely determined by the local-field distribution or equivalently, the dynamical structure 
factor of the Ising bath. We present analytic results for the Ising spin chain bath at arbitrary 
temperature for different distributions of the intrabath interaction strengths. We find clear evidence 
of non-Markovian behavior in the low temperature regime. We also consider baths described by Ising models 
on higher-dimensional lattices. We find that interactions lead to a significant reduction of the decoherence.
An important feature of interacting spinbaths is the saturation of the asymptotic 
Markovian decay rate at high temperatures, as opposed to the conventional Ohmic boson bath. 
 
\end{abstract}
\pacs{}

\maketitle

\section{Introduction.}

Recent developments in nanophysics have made possible the  use of  the charge and spin dynamics
of electrons to develop new technologies like spintronics and  quantronics.  This has moreover led to the possibility of
using the electron spin, or other more complex entities like the phase in  Josephon junctions 
to fabricate qubits  for quantum computing.  However,
the utility  of these nanosystems as qubits is strongly limited by
their coupling to  the  omnipresent dissipative environment. The environment destroys the coherence of the  qubit over a certain time
scale and a  lot of recent  theoretical and experimental activity has
focused on ways and means to increase this time scale.
This  clearly emphasizes the importance of understanding the effects
engendered by the coupling of a two level system to a dissipative bath.

The fact  that the environment plays a crucial role in the
physics of small quantum systems has been well known since the pioneering
work of Ref.{\onlinecite{Caldeira-Leggett}}, where it was shown that the coupling of a
two level system  to an Ohmic boson bath could
effectively suppress the tunneling  of the two level system.  In the context of decoherence,   
the most commonly studied problem is the spin-boson model \cite{QDS} which describes the effect 
of a dissipative bosonic bath on a central spin, where the spin can be an effective description 
of a system whose discrete lowest energy levels dominate the physics at low enough temperatures 
and the bosons are often the phonons present in the system. A physical manifestation of the spin boson 
problem is a nanomagnet (described by a giant spin) coupled to phonons \cite{stamp-review}.  
However, for many  practical realisations of a
central spin or a qubit (spin $1/2$), a spin bath comprising  other spins 
might be the principal source of decoherence. This is indeed the case in semiconducting 
quantum dots,    where the nuclei with non zero spins constitute the spin bath and interact with 
the central  electronic spin  in the dot  via the hyperfine interaction \cite{Coish-Loss,Marcus}. 
Another manifestation of a spinbath occurs in  Si:P \cite{Hu}. 
The abundance of spin baths in real systems, necessitates an understanding of their effect on decoherence. 
Unlike the case of bosonic baths often
modeled as a collection of harmonic oscillators,   spin baths  can exhibit a wide range of phenomena
depending on the interactions between the spins, residual anisotropies etc.  Clearly one expects any resulting
decoherence of the central spin to depend rather crucially on the
underlying nature of the spin bath and its coupling to the central spin. 

Earlier studies which considered independent spins in the bath seemed to indicate that spin baths 
were not qualitatively different from bosonic baths \cite{Hanggi,Hanggi1}.
More interestingly,  recent studies of decoherence  induced by spin baths
described by mean field Hamiltonians have demonstrated that
interactions between the bath spins can be used as a lever to augment
 the time scales over which the system decoheres
\cite{italian-meanfield, sib,Yuan}. These results were however
obtained either numerically for a bath with a small number of spins or
 for the special case where the bath Hamiltonian commutes with the bath-central spin 
coupling Hamiltonian leading  to an effective classical decoherence.
 A more robust treatment of  intrabath interactions  was presented in 
 Ref.\onlinecite{cb}  where the authors   studied numerically the  zero temperature {\it quantum }decoherence of two coupled spins engendered by a  bath described by the random transverse Ising model.  
The authors used this model to argue that  the central spin decoheres differently, depending 
on whether the spinbath has a  regular or chaotic   spectra.   
Despite their various drawbacks, these works  collectively  highlight the
importance of interactions and disorder  in the bath. 
Moreover, disordered spin baths warrant further attention because
both interactions (often dipolar) and disorder are   present in real spin bath  systems like
quantum dots in  semiconducting heterojunctions and in Si:P.

In this paper, we re-examine the decoherence
induced by disordered interacting spinbaths at finite
temperatures. More precisely, we study the  effect of an Ising 
bath with random spin-spin interactions  on the coherence of a
single spin $1/2$.  The random interactions are
characterized by their variance $\Delta^2$  where  $\Delta$ is analogous 
to the cut off frequency for a bosonic bath as well as a mean value $J_0$ 
which has no bosonic counterpart.  Our choice of an Ising bath is primarily to
facilitate  an analytical study of the problem at finite temperatures in the
thermodynamic limit. To ensure a quantum decoherence of the central
spin in our model,  the central spin is coupled to the transverse spin
components of the bath. To better comprehend the effect of the bath, we consider 
a model in which the time evolution of the central spin is exclusively governed 
by its coupling to the bath. For such a model, the problem is exactly soluble 
for a bath comprising independent spins/bosons. However, when intra-bath interactions are 
present, no exact solution can be obtained and one has to take recourse to approximate 
methods. In this paper, we only study the limit of a
weak coupling of the central spin to the bath, where robust analytical methods are
available to study the problem in an unbiased manner. 
We use the resolvent operator method, which takes us beyond
the oft used Markovian master equation approaches to study the decoherence  of the central
spin induced by a perturbative coupling to the Ising spin bath. 
An interesting aspect of  our work is that  in the absence of any dynamics intrinsic to the central spin, 
for weak coupling to the bath, the decoherence is primarily dictated by the local-field distribution 
 or equivalently, the dynamic structure factor of the bath spins. For a bath described by an Ising spin chain,
 we obtain the Markovian decoherence time scale  and the non Markovian corrections   as a function of
the temperature and  the parameters of the bath.
We  also discuss the case of Ising spins on various lattices in the
high temperature regime and the case of the infinite-ranged Sherrington-Kirkpatrick
spin glass model.  

The paper is organized as follows: we present the model and
derive a general expression for the decoherence of the central spin in \sref{sec:model} followed by
a  discussion of the weak coupling regime in \sref{sec:wcr}. 
We then present our results for different disordered Ising spin baths in \sref{sec:DIsb}.

\section{Model}\label{sec:model}

 We present  the model used to  study the decoherence of  the central spin 
$\sigmaV_c$, weakly coupled 
to a bond-disordered bath of $N$ Ising spins $\sigmaV_i$ in the thermodynamic limit $N\to\infty$.
The total Hamiltonian  describing the combined system of the central spin and the spin bath is
given by 
\begin{eqnarray}
H&=&H_B+\sigma^x_cV \\
&\equiv&-\sum_{(ij)} J_{ij} \sigma_i^z \sigma_j^z -\sigma^x_c \sum_{i} \lambda_{i} \sigma_i^x 
 \label{H}
\end{eqnarray}
where,
$\sigma^x_c$ is the $x$-component Pauli operator of the central spin and  $\sigma_i^x$ and $\sigma_i^z$ 
denote the Pauli operators of the bath spins and $J_{ij}$ are the
interaction strengths between the bath spins. 
 Depending on the details of the model studied, $(ij)$ could represent
interactions between nearest neighbour spins or interactions of infinite range.
 In contrast to the models studied in Ref.~\onlinecite{italian-meanfield}, where the bath hamiltonian $H_B$ and the
bath operator 
$V$ that couples to the central spin  commute,  here our choice of $V$ is such that   $[H_B, V] \neq 0$.
The  central spin and spin bath  coupling   is characterized by the parameters $\lambda_i$. 
Since, we are interested in the influence of disorder as well as the  tendency of the
system to order, we consider random  interaction  energies $J_{ij}$ which are 
quenched random variables drawn from a distribution $p(J)$ with mean 
$J_0$ and variance $\Delta^2$.  
Though  the central spin  does not have any intrinsic dynamics, its coupling
to the bath generates a non trivial dynamical behaviour. We note that $H_B$ is the usual
Ising spin glass Hamiltonian which has been well studied in the past\cite{spinglassbook}. Depending on the
distribution of the spin-spin interactions and the dimensionality, this model  can exhibit  ferromagnetic, antiferromagnetic   or even spin glass order in some temperature range.  Since these phenomena have
ramificiations for the collective behaviour of the bath,  it is reasonable to expect  the
resulting decoherence to depend crucially on the underlying order in the bath.

 It is important to note that the formalism developed in this section and \sref{sec:wcr}  is  {\it a priori}
 applicable to any  
 bath hamiltonian $H_B$ (bosonic baths, Heisenberg spin baths, baths with both spins and bosons etc). 
  For a Hamiltonian of the form \eqref{H},  since $\sigma^x_c$ is a constant of motion,  it is
convenient to directly study  the time  evolution of the reduced
density matrix of the 
central spin 
 \begin{equation}
\rho(t)=\mathrm{Tr}_B \left( e^{-iHt} \Omega e^{iHt} \right)
\end{equation}
where $\Omega$ is the initial density matrix of the composite system consisting of the central 
spin and the bath and $\mathrm{Tr}_B$ denotes the partial trace over the bath degrees of freedom.  We use  the units $\hbar=k_B=1$ in this paper.
 Denoting  the eigenstates of $\sigma^x_c$ by $|\! \gets \rangle$ and $|\! \to \rangle$, we see that due to the property of the density matrix $\mathrm{Tr} [\rho(t)] = 1$  and the
 stationarity of $\mathrm{Tr} [\rho(t) 
\sigma^x_c]$,   the diagonal elements of the density matrix given by the 
 populations $\langle \gets \! | \rho(t) | \! \gets \rangle$ 
and $\langle \to \! | \rho(t) | \! \to \rangle$ remain
 constant.  Only 
the coherences $\langle \to \!|  \rho(t) | \! \gets \rangle=\langle \gets  \!| \rho(t) | \! \to \rangle^*$  representing the off-diagonal elements of 
the density matrix $\rho(t)$ change with time. Often these coherences vanish  at long times resulting in a decoherence of the central
spin  i.e., 
the asymptotic state of the central spin is a mixture of the states $|\! \gets \rangle$ and $|\! \to \rangle$ 
irrespective of the nature of the initial state $\Omega$.

To simplify the calculation,  we assume that at time $t=0$, the central spin and the bath are 
disentangled resulting in a factorizable   initial density matrix: $\Omega=\rho(0)  \otimes \rho_B$. 
We further suppose that initially the central spin is in a pure state 
$ |\psi \rangle = \alpha|\! \gets \rangle + \beta|\! \to \rangle$ ($\rho(0) = |\psi \rangle \langle \psi |$) 
and that the bath is at thermal equilibrium with temperature $T\equiv 1/\beta$\,:
\begin{equation}
\rho_B = \frac{e^{-\beta H_B}}{Z} \label{rhoB}
\end{equation} 
where $Z=\mathrm{Tr} \, \exp(-\beta H_B)$ is the bath partition function. 
With these initial conditions we obtain the time evolved reduced density matrix 
\begin{widetext}
\begin{equation}
\rho(t)= |\alpha|^2 \vert \!\gets \rangle \langle \gets \! \vert + 
|\beta|^2 \vert \! \to \rangle \langle \to \! \vert + 
M(t) \alpha^* \beta \vert \! \to \rangle \langle \gets \! \vert 
+ M(t)^* \alpha \beta^* \vert \! \gets \rangle \langle \to \! \vert
\end{equation}
\end{widetext}
where the factor
\begin{equation}
M(t)= \mathrm{Tr} \left( e^{-i(H_B+V)t} \;\rho_B\; e^{i(H_B-V)t} \right) \label{Mt}
\end{equation}
is a measure of the decoherence induced by the bath at time $t$. 
Here $\mathrm{Tr}$ denotes the usual trace as $H_B$ and $V$ are operators in the bath Hilbert space. 
Under a rotation of $\pi$ around the $z$-axis,
while  $H_B$ and $\rho_B$ remain unchanged,  $V \to -V$. Consequently the coherence $M(t)$ is a real number.
As mentioned in the introduction, though the coherence $M$ is easy to evaluate for baths consisting of
non-interacting spins/bosons (see \aref{Isb}),  it is rather difficult to estimate for an interacting spin bath
  for arbitrary values of the coupling to
the central spin.
We remark that $M(t)$ is  related to Loschmidt echoes of the bath which characterize 
its sensitivity to perturbations in its equations of motion \cite{LE,LE1,fazio}.

\section{Weak coupling regime}\label{sec:wcr}

In this section, we present a perturbative formalism to calculate $M(t)$  valid  for  weak
coupling  to the bath i.e., the energy scale of the operator $V$ is much smaller than all the energy scales of the bath.  
We use the resolvent operator method which goes beyond the Born-Markov approach  and, though
perturbative, is intrinsically capable of handling non-Markovian time evolutions.  We obtain a tractable expression
for the decoherence in the weak coupling regime.

\subsection{Resolvent operator method}

To determine the complete time evolution of the coherence $M$ in the weak coupling limit $V \to 0$, we first express it 
in terms of  a {\it self energy}  $\Sigma$\cite{ROM,ROM1}.  To obtain   the self energy, it is
convenient to work with the  Laplace transform  of $M(t)$ 
\begin{equation}
{\tilde M}(z)= -i \int_0^\infty dt \, e^{izt} M(t)  \label{LTdef} 
\end{equation}  
where  $z$ is a complex variable with $\mathrm{Im} z >0$. 
As shown in \aref{Dse}, this Laplace transform can be written as 
\begin{equation}
{\tilde M}(z)= \left[ z-\Sigma(z) \right]^{-1} \label{LT} 
\end{equation}  
\noindent
where the self-energy $\Sigma$ is given by 
\begin{equation} 
\Sigma(z)= \mathrm{Tr} \Big[ {\cal L}_V \rho_B + {\cal L}_V {\cal Q} (z-{\cal Q}\, {\cal L} \,{\cal Q})^{-1}
{\cal Q} {\cal L}_V \rho_B \Big] \label{Sigma}
\end{equation}
In the above expression, the superoperators  ${\cal Q}$, ${\cal L}_B$, ${\cal L}_V$ and ${\cal L}$ are defined by their  actions on  any operator $A$ acting on the bath Hilbert space:
${\cal Q} A  =  A - \mathrm{Tr} (A) \rho_B$, ${\cal L}_B A = [H_B, A]$, ${\cal L}_V A = V A + A V$  and ${\cal L}={\cal L}_B+{\cal L}_V$. 
As $\rho_B$ is a density matrix, ${\cal Q}$ is a projection operator, i.e. ${\cal Q}^2={\cal Q}$. 
Note that ${\cal L}_V$ is not a Liouville operator whereas ${\cal L}_B$ is the Liouville operator 
corresponding to the bath Hamiltonian $H_B$.  We reiterate that  the above derivation for \eqref{LT} and \eqref{Sigma}  is independent of the specific 
Hamiltonian $H_B$ and coupling operator $V$ considered in this paper. 
The self-energy $\Sigma$ can  now be expanded perturbatively in the interaction operator $V$. 
The second order result  is 
 given by the expression \eqref{Sigma} with ${\cal L}$ replaced 
by the bath Liouvillian ${\cal L}_B$.
The first-order term $ \mathrm{Tr} ( {\cal L}_V \rho_B )=2\mathrm{Tr} (V \rho_B )$ vanishes for the Ising 
spin bath Hamiltonian  $H_B$ and the interaction operator $V$ defined by \eqref{H}. 
Therefore, the first non-zero  contribution to the self-energy 
is given by the second-order term $\Sigma_2$ which can be rewritten in terms of the time-dependent 
symmetrised correlation function of $V$ (see \aref{Sose}) :
\begin{equation}
\Sigma_2(z) = -2i \int_0^\infty dt\, e^{izt} \left[ \langle V(t)V\rangle + \langle VV(t)\rangle \right] \label{soSigma}
\end{equation}       
where $\langle...\rangle$ refers to the thermal average over bath spin configurations. Neglecting higher order contributions to $\Sigma$ in \eqref{LT} is equivalent to the Born approximation \cite{QDS}. 
We will see in the following that this approximation can describe the   decoherence at all timescales whereas 
a direct expansion of the coherence $M$ gives only the short-time evolution. 
We remark that in the case of Heisenberg spins, an underlying magnetic order could result in a first order 
contribution to the self energy which would then lead to an oscillatory behaviour of $M(t)$.

The advantage of the resolvent operator formalism is that  we can use the analyticity properties of the self 
energy $\Sigma$ to obtain a
tractable expression for  the coherence $M$. 
As shown in \aref{Dse}, since $H_B$ and $V$ are Hermitian operators,
the spectrum of the operator ${\cal L}$ 
is real and hence the self-energy $\Sigma$ is analytic in the upper (lower) half plane. 
Furthermore, since  the spectrum of $\cal L$ in the thermodynamic limit  is expected to be a continuum
  for the models considered in this paper, the self-energy $\Sigma$ manifests   a branch cut on the real axis.  
The coherence $M$ can thus be written 
in terms of the real functions $\Lambda$ and  $\Gamma$ defined by
\begin{equation}
\Lambda (E)-i\Gamma(E)= \lim_{\eta {\to} 0^+} \Sigma(E+i\eta) \label{DeltaGamma}
\end{equation} 
where $E$ is real. Since $M(t)$ is real, the functions $\Lambda$ 
and $\Gamma$ satisfy $\Lambda(-E)=-\Lambda(E)$ and $\Gamma(-E)=\Gamma(E)$. 
Performing the inverse Laplace transform of \eqref{LT} and 
taking the limit $\eta {\to} 0^+$, we obtain 
 \begin{equation}
\Theta(t) M(t) = \frac{i}{2\pi} \int dE \frac{e^{-itE}}{E-\Lambda(E)+i\Gamma(E)} \label{ThetaM} 
\end{equation}     
where $\Theta(t)$ is the Heaviside step function. Moreover, 
integrating $\Sigma(z)/(z-E)$ along an appropriate contour in the upper half plane, one obtains the Kramers-Kronig 
like dispersion relation 
 \begin{equation}
\Lambda(E)-i\Gamma(E) = -\frac{i}{\pi} {\cal P} \int \frac{dE'}{E'-E} \left[ \Lambda(E')-i\Gamma(E') \right]  
\label{Ht}
\end{equation}   
where ${\cal P}$ denotes the Cauchy principal value. This shows that $\Theta(t) M(t)$ 
is exclusively  determined by $\Gamma$ (or $\Lambda$).

\subsection{Non-Markovian evolution }\label{sec:NME}

In this section, we analyze \eqref{ThetaM} to determine  $M(t)$ at any time $t$ in the weak coupling regime. 
We first note that differentiating \eqref{ThetaM} with respect to the time $t$ yields the conditions
 $M(0)=1$ and $\partial_t M(0)=0$.
Differentiating  \eqref{ThetaM} further
  and taking 
the limit $V\to 0$, we find,  to second order in $V$,   
\begin{eqnarray}
\Theta(t) \partial^2_t M & \simeq &- \frac{1}{2\pi} \int dE e^{-iEt} [ \Gamma_2(E) + i\Lambda_2(E) ]  \nonumber \\
 & \simeq &- \frac{1}{\pi} \Theta(t) \int dE \, e^{-iEt} \, \Gamma_2(E) 
\end{eqnarray}        
where $\Gamma_2$ and $\Lambda_2$ denote the second-order terms of the functions $\Gamma$ and $\Lambda$.  Note that to obtain the second equality we have used the relation \eqref{Ht} and the Fourier 
transform of $2\Theta(t)-1$. 
A solution to the above differential equation, for $t > 0$, is 
\begin{equation}
M(t) \simeq 1 - \frac{2}{\pi} \int dE\,  \frac{\sin (tE/2)^2}{E^2} \Gamma_2(E) \label{Mapproxst} 
\end{equation} 
where we have taken into account the symmetry of the function $\Gamma_2$. Though this   Fermi golden rule like 
approximation   yields  the correct short time behaviour,  it leads to the false result
 $M(t) \simeq -\Gamma_2(0)t$ for $t \to \infty$  
and hence is  invalid for an arbitrary time $t$.

To obtain the long-time decoherence, we evaluate the integral 
\eqref{ThetaM} using the analytic continuation of $\Sigma$ from the upper half plane 
to the lower half plane (second Riemann sheet). If the functions $\Gamma$ and $\Lambda$ 
are analytic in the vicinity of $E=0$, the analytic continuation of $\Sigma$ in 
the second Riemann sheet is given by 
\begin{equation} 
{\tilde \Sigma} (z) = - i\Gamma(0) + \Lambda'(0) z + O(|z|^2)
\end{equation}  
for small $z$ where $\Lambda'$ denotes the derivative of the function $\Lambda$ 
with respect to $E$. Note that the symmetry properties of the functions $\Lambda$  and $\Gamma$  
ensure $\Lambda(0)=0$ and $\Gamma'(0)=0$. 
The coherence \eqref{ThetaM} is  principally determined by the singularities of 
$[z-{\tilde \Sigma} (z)]^{-1}$, one of which is a pole at $z_0 = - i\Gamma_2(0) + O(V^4)$ 
close to the real axis. Every other singularity lies beyond a finite distance $\epsilon^{-1}$
from the real axis   determined by the temperature and 
the bath parameters. Consequently, for times 
$t \gg \epsilon$, the residue $\exp(-iz_0t)[1-\partial_z {\tilde \Sigma} (z_0)]^{-1}$ 
dominates and thus  in the weak coupling limit $V \to 0$,
\begin{equation} 
\ln M(t) \simeq  - \Gamma_2(0)t + \Lambda_2'(0) \quad . \label{Mapproxlt}
\end{equation}  
 Combining the approximations for short times   and long times given by \eqref{Mapproxst} and  \eqref{Mapproxlt} 
 respectively, we see that
 in the weak coupling regime the  decoherence  at any time $t$  is described by
\begin{equation}
\ln M(t) \simeq - \frac{2}{\pi} \int dE\,  \frac{\sin (tE/2)^2}{E^2} \Gamma_2(E) \label{Mapprox} \quad .
\end{equation} 
This equation shows that  $M(t)$ in the weak coupling regime is determined by the  entire function $\Gamma_2$.
For asymptotic times, we see from \eqref{Mapprox} that the coherence of the central spin is essentially given by $M(t) \simeq \exp[-\Gamma_2(0)t]$
which 
 is simply the solution of the Markovian master equation obtained 
within the Born-Markov approximation \cite{QDS}. 
However, for the  short and intermediate time evolution of $M$,  the   full energy dependence of   $\Gamma_2$ comes into play
which can then  lead to a {\it non-exponential decay } i.e., non-Markovian behaviour of the coherence. Depending on the temperature and
bath parameters, the asymptotic
Markovian regime could even disappear, provided that  $\Lambda_2'(0)$ goes to infinity. We remark that Eq.\eqref{Mapprox} 
is valid to all orders in $V$ for a bath of independent bosons.

\section{Disordered Ising spin baths}\label{sec:DIsb}

In this section, we use the formalism of the previous section to determine the decoherence induced by 
various disordered Ising spin baths. Though the spinbath models of real systems are expected to be more 
complicated than the Ising Hamiltonians considered here, we nonetheless study these systems in various 
dimensions to understand the effect of these simpler systems on the decoherence. 
More precisely, we study the effect of the disordered Ising spin chain on the coherence
$M$ of the central spin as a function of the temperature and the bath parameters $J_0$ and  $\Delta$. 
We also make predictions for the infinite-ranged Ising model also known as the Sherrington-Kirkpatrick 
model and for  Ising models  in higher dimensions in the high temperature regime. Before we embark on 
in-depth calculations of $M$, we show that for any Ising bath, the decoherence of the central 
spin in the weak coupling regime is intimately linked to the local-field distribution of the bath.

\subsection{Local-field distribution}

As shown in \sref{sec:wcr}, the decoherence in the weak coupling regime is determined by the time-dependent 
correlation function $\mathrm{Re} \langle V(t) V \rangle$, through \eqref{Mapprox} and \eqref{soSigma}. 
For the case of the Ising bath Hamiltonian defined in (\ref{H}),  this correlation
can be obtained  from the local-field distribution of the bath. To see this,
it is useful to work in the eigenbasis  $|\{\sigma_i\} \rangle$  of $H_B$ ~: $H_B|\{\sigma_i\} \rangle=-
\sum_{(ij)} J_{ij} \sigma_i \sigma_j|\{\sigma_i\} \rangle$. 
As the matrix element $\langle \{\sigma_i\} | \sigma_k^x  |\{\sigma'_i\} \rangle$ is nonvanishing only for spin configurations 
$\{\sigma_i\}$ and  $\{\sigma'_i\}$ where $\sigma'_i=\sigma_i(1-2\delta_{ik})$, the time-dependent correlation of $V$ is a sum 
of local correlations :  $\langle V(t) V \rangle=\sum_k \lambda_k^2 \langle \sigma_k^x(t) \sigma_k^x  \rangle$. We find
\begin{equation}
\mathrm{Re} \langle V(t)V\rangle = \frac{1}{Z} \sum_{k,\{\sigma_i\}} \lambda_k^2  e^{\beta \sum_{(i j)} J_{ij} \sigma_i \sigma_j}  
\cos \left( 2t \sum_{i}\nolimits^{(k)}J_{ki}\sigma_i\right) \label{VV1}
\end{equation}    
where $\sum_{i}^{(k)}$ denotes a sum over the spins $\sigmaV_i$ interacting with the spin $\sigmaV_k$. 
Note that the term  $\sum_{i}^{(k)} J_{ki} \sigma_i$  in \eqref{VV1} is the effective local field  acting on the spin at site $k$ 
generated by the configuration $\{\sigma_i\}$. Since $\mathrm{Re} \langle \sigma^x(t) \sigma^x  \rangle=\cos(2th)$ for an isolated spin {\boldmath $\sigma$} 
in a field $h$ parallel to the $z$-axis, we rewrite \eqref{VV1} as    
\begin{equation}
\mathrm{Re} \langle V(t)V\rangle  =  \sum_k \lambda_k^2 \int dh\, P_k(h) \cos(2th) 
\end{equation}   
where $P_k(h)= \left\langle \delta \left( h- \sum_{i}^{(k)} J_{ki} \sigma^z_i \right) \right\rangle$ can be interpreted as the distribution of 
the local field $h$ at site $k$ at temperature $T$. This interpretation 
is also valid for the bath thermodynamic quantities such as magnetisation or specific heat \cite{lfd}. 

It is now rather straightforward to obtain  the function $\Gamma_2$ which is the
crucial ingredient to determine the coherence $M$ in the weak coupling regime. To avoid computational
complexity, in the rest of the paper, 
 we consider equal couplings $\lambda_k=\lambda N^{-1/2}$.  With this choice, the function $\Gamma_2$ reads 
\begin{equation}
\Gamma_2(E) = 2\pi  \lambda^2 P (E/2)   \label{GammaP}
\end{equation}  
where the local-field distribution $P$ is the spatial average 
\begin{equation}
P(h)=\frac{1}{N} \sum_k P_k(h) =  \frac{1}{N} \sum_k \left\langle \delta \left( h- \sum_{i}\nolimits^{(k)} J_{ki} \sigma^z_i \right) \right\rangle \quad .
\label{lfd}
\end{equation}
\noindent
The thermodynamic limit of this expression is unambiguously defined and $P$ is self-averaging \cite{salfd} 
i.e., in the limit $N \to \infty$, $P$ is given by 
 bond disorder  average of any distribution $P_k$. For an Ising spin system, the local-field 
distribution $P$ is temperature dependent and determines the thermodynamic quantities 
and the dynamic linear response of the system \cite{lfd,Plefka}. Eq. (\ref{GammaP}) shows that the local-field 
distribution is also an important characteristic of an Ising system considered as a bath.

We remark that for a bath of independent spins \cite{Hanggi}, i.e. $H_B=-\sum_i h_i \sigma_i^z$ in \eqref{H}, 
the function $\Gamma_2$ is given by \eqref{GammaP} with the field distribution 
$P(h)=\sum_k (\lambda_k/\lambda)^2 \delta(h-h_k)$. 
In this case, since the spins are non-interacting,  the distribution $P$ is arbitrary and temperature independent
and the ensuing 
 decoherence of the central spin is  temperature independent. 
This feature of a temperature independent decoherence  induced by a bath of  independent spins 
seen in the weak coupling limit 
is also seen in  the exact non perturbative result for the coherence $M(t)$ obtained in \aref{Isb}.

\subsection{Ising spin chain}
In this section we consider a 1D Ising bath described by the Hamiltonian 
\begin{equation}
H_B =-\sum_{i=1}^{N-1} J_{i} \sigma_i^z \sigma_{i+1}^z \quad . \label{1DH}
\end{equation}
The spin at site $i$ interacts with its nearest neighbors with interaction strengths $J_i$ and $J_{i-1}$. 
The $J_i$ are a quenched set of random bonds drawn independently from a distribution $p$ with mean 
$J_0$ and variance $\Delta^2$. 

\subsubsection{Born self-energy}

\begin{figure}

\centering \includegraphics[width=0.45\textwidth]{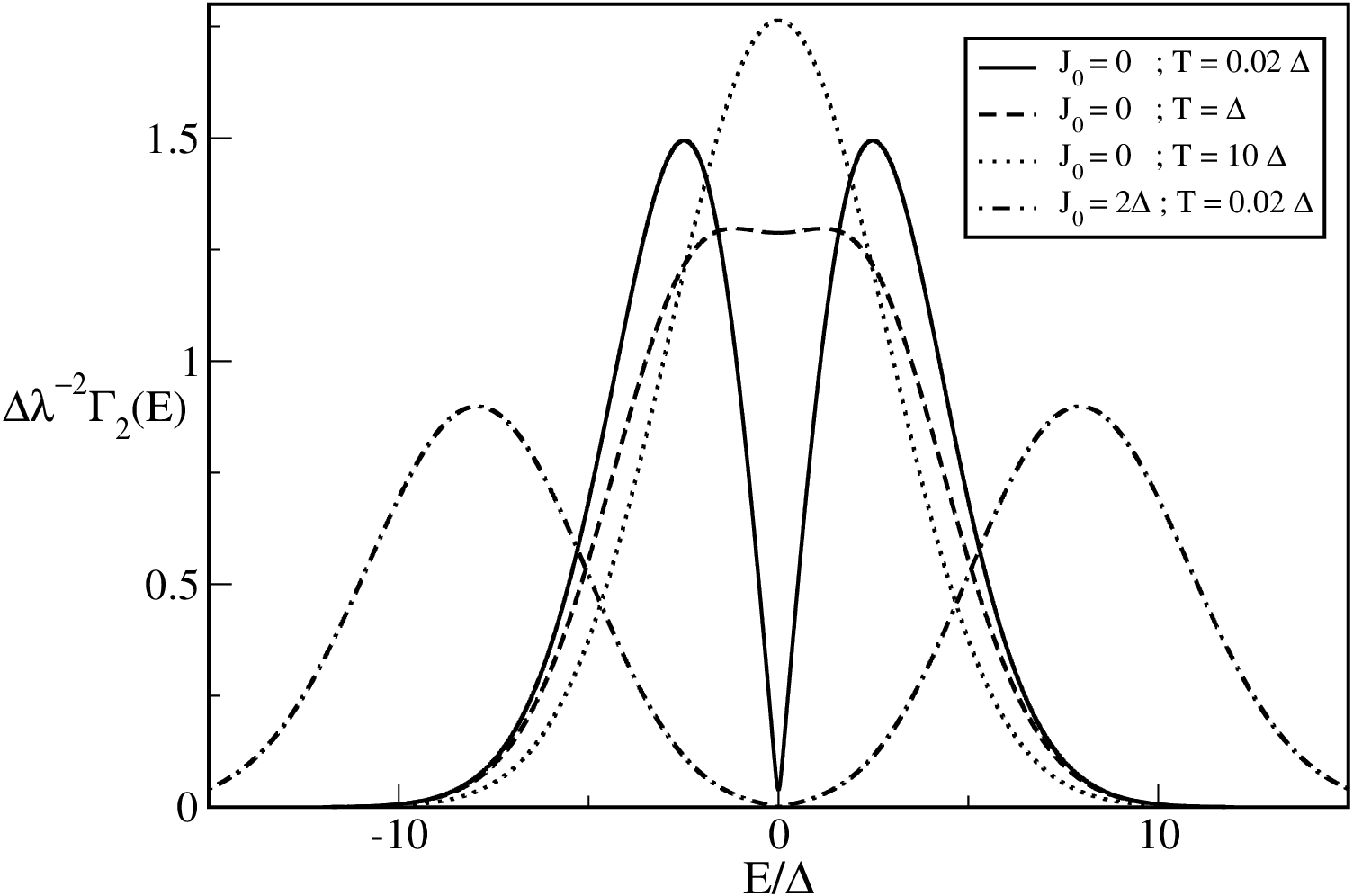}

\caption{\label{fig:gamma} The real part of the dimensionless second-order self-energy
$\Gamma_2 \Delta / \lambda^2$
as a function of $E/\Delta $ for a Gaussian bond distribution
with mean values $J_0=0$, $2\Delta$ and temperatures $T=0.02\Delta$, $\Delta$, $10\Delta$.}
\end{figure}

To obtain the coherence $M$ we need the time-dependent correlation of 
the coupling operator $V=-\sum_i \lambda_i \sigma_i^x$. As shown earlier, this correlation 
is given by $\mathrm{Re} \langle V(t) V \rangle=\sum_k \lambda_k^2 \mathrm{Re} \langle \sigma_k^x(t) \sigma_k^x  \rangle$.
Here the time dependent spin-spin correlations can be written in terms of the static correlation
function as 
\begin{eqnarray}
\mathrm{Re} \langle \sigma_k^x (t) \sigma_{k}^x \rangle &= &\cos(2tJ_{k-1}) \cos(2tJ_{k})  \\ &&
- \langle \sigma_{k+1}^z \sigma_{k-1}^z \rangle \sin(2tJ_{k-1}) \sin(2tJ_{k}) \nonumber
\end{eqnarray}
where $\langle \sigma_{k+1}^z \sigma_{k-1}^z \rangle=\tanh(\beta J_{k-1}) \tanh(\beta J_{k})$ is related 
to the derivative of the partition function 
\begin{equation}
Z=2^N \prod_{i=1}^{N-1} \cosh (\beta J_{i})
\end{equation}
with respect to the interaction strengths $J_{k}$ and $J_{k-1}$. 
The choice $\lambda_k=\lambda N^{-1/2}$ yields a variance of 
$\mathrm{Re} \langle V(t)V\rangle$ of order $N^{-1}$ and a mean of order $N^0$.
Consequently,  $\mathrm{Re} \langle V(t)V\rangle$ and hence $\Sigma_2$ are self-averaging 
in the thermodynamic limit. The second-order self-energy is thus given by the average of \eqref{soSigma} 
over bond disorder. We obtain 
\begin{widetext}
\begin{equation}\label{Gamma1D}  
\Gamma_2(E)=2\pi \lambda^2 \int dJ \, p_e (J) p_e (J+E/2) \left[ 1 - \tanh(\beta J)\tanh(\beta J+\beta E/2) \right]  
\end{equation}
\end{widetext}
where $p_e (J)=[p(J)+p(-J)]/2$ is the symmetrized bond distribution. 
Note that the integral on the right side is the local-field distribution \eqref{lfd} 
for the spin chain\cite{salfd,1Dlfd}. Equation \eqref{Gamma1D} is valid for all  bond distributions $p$. 
Moreover, for distributions $p$ symmetric around their mean values, 
as $\Gamma_2$ depends only on the symmetrized distribution $p_e$ the self-energy $\Sigma_2$ is the same for opposite 
means $\pm J_0$. This shows that, in this case, though the decoherence is influenced by the interactions in the bath, 
it is  insensitive to the
underlying ferromagnetic or antiferromagnetic nature of the interactions. This feature can be understood as follows. 
The spin chain Hamiltonian \eqref{1DH} and the local field operator 
$J_{k-1} \sigma_{k-1}^z + J_{k} \sigma_{k+1}^z$ are invariant under 
the transformation $[J_{i},\sigma^z_i] \to [-J_{i},(-1)^{i-k} \sigma^z_i]$. 
Consequently, the local-field distribution $P$ and hence 
the function $\Gamma_2$ are invariant 
under the transformation $J_0 \to -J_0$. To complete the determination of the second-order self-energy, its real part 
$\Lambda_2$ can be obtained from the function $\Gamma_2$  using \eqref{Ht}.  
In \fref{fig:gamma}, we plot
$\Gamma_2$ for different values of $J_0$ and the temperature $T$.

\subsubsection{Markovian evolution}

As shown in \sref{sec:wcr}, in the weak coupling limit $\lambda \to 0$, the evolution of 
the coherence $M$ is essentially Markovian : $M(t) \simeq \exp(-\gamma t)$.  For the 
disordered Ising chain, the  decoherence 
rate $\gamma$ is given, to lowest order in $\lambda$, by
\begin{equation}
\gamma=\Gamma_2(0)=2\pi \lambda^2 \int dJ \, p_e (J)^2 \left[ 1 - \tanh(\beta J)^2 \right] \quad . \label{rate}   
\end{equation}   
The rate $\gamma$ increases monotonically from $\gamma=0$  at $T=0$  and saturates to the value
\begin{equation}
\gamma_\infty = 2\pi \lambda^2 \int dJ \, p_e (J)^2  \label{ratehT}   
\end{equation}    
as $T \to \infty$. At low temperatures,    $\gamma \simeq 4\pi\lambda^2 p (0)^2\,T$  for disorder distributions  with $p(0) \neq 0$. 
The vanishing of $\gamma$ 
at zero temperature can be understood as follows \cite{salfd}. At $T=0$, the local
field distribution is completely dictated by the spin structure of the  ground state 
$| \{ \sigma_i \} \rangle$ of the bath Hamiltonian \eqref{1DH}.   Since $J_i  \sigma_i \sigma_{i+1} > 0$ for any pair 
of neighboring spins in the ground state, this inevitably leads  to nonvanishing local fields 
$J_{i-1}  \sigma_{i-1} + J_i \sigma_{i+1} = (|J_{i-1}| + |J_i|) \sigma_i$. This implies $P(h=0)=0$  and hence $\gamma=2\pi\lambda^2 p(0)=0$. 

\begin{figure}
\centering \includegraphics[width=0.45\textwidth]{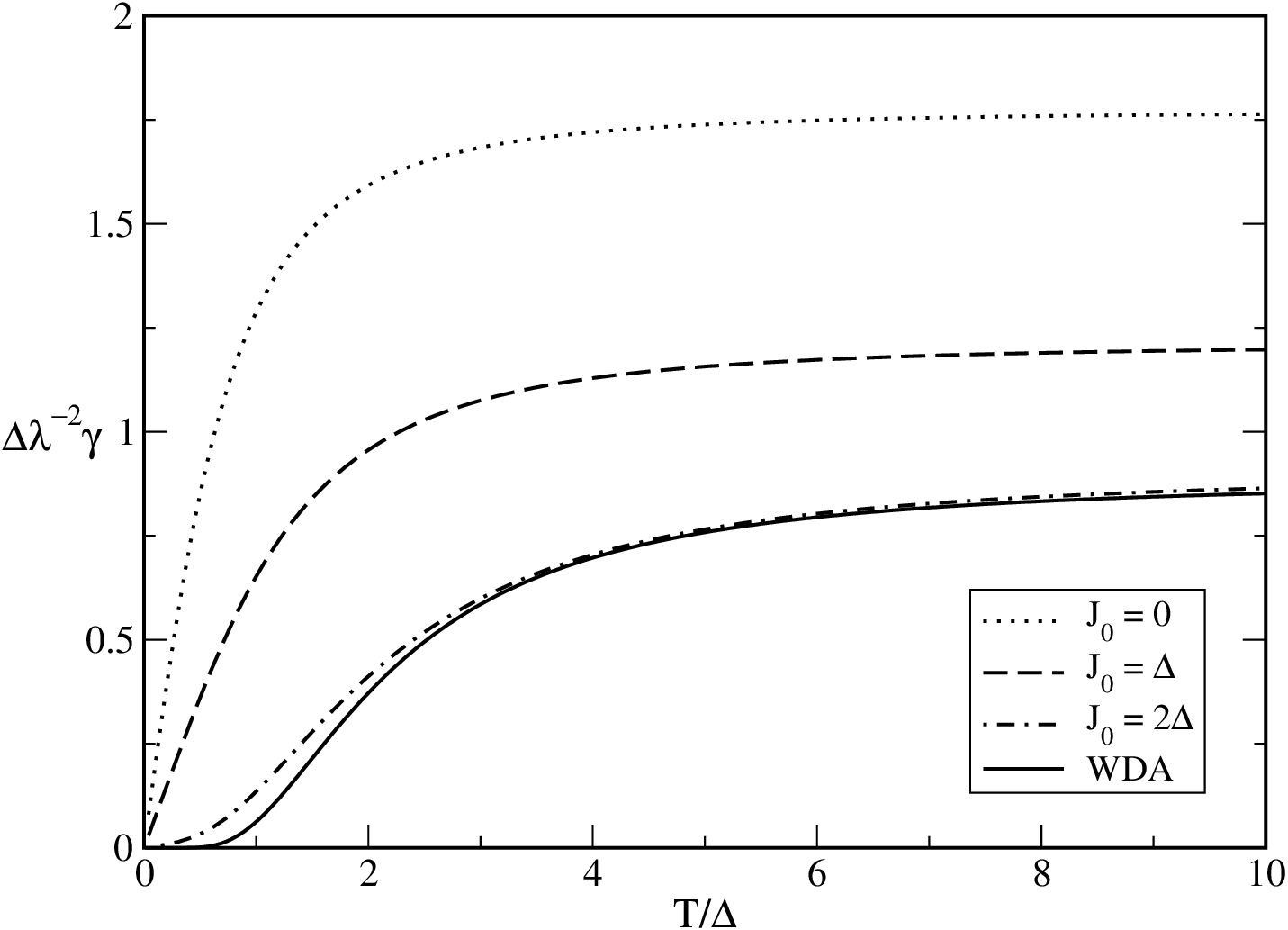}

\caption{Dimensionless rate $\gamma \Delta/\lambda^2$ as a function of the temperature $T/ \Delta$ 
 for a Gaussian bond distribution with mean $J_0 = 0$, $\Delta$ and  $2\Delta$. The weak disorder 
approximation \eqref{wdapprox} is shown for $J_0 = 2\Delta$.}
\label{fig:rate}

\end{figure}

We now study  the influence of  the bond distribution   on the rate $\gamma$. 
In the weak disorder regime $\Delta  \ll |J_0|$, 
the distribution $p_e(J)^2$ practically vanishes for $J\neq \pm J_0$ thus the rate $\gamma$ is given by 
\begin{equation}
\gamma \simeq \pi \lambda^2 \int \! dJ \, p(J)^2 \;\;  \left[ 1 - \tanh(\beta J_0)^2 \right] \quad . \label{wdapprox}  
\end{equation}   
In this regime, the temperature variations of $\gamma$ are independent of the form of the bond distribution 
$p$ and are exclusively determined by the mean value $J_0$. 
In  \fref{fig:rate}, we plot  the temperature dependence of  $\gamma$ 
for various values of the mean 
interaction strength $J_0$ in the case of a Gaussian bond distribution
\begin{equation} \label{Gbd}
p(J)=\frac{e^{-(J-J_0)^2/2\Delta^2}}{\sqrt{2\pi} \, \Delta } \quad . 
\end{equation}    
The linear regime at low temperature given by $\gamma \simeq 2(\lambda/\Delta)^2 \exp(-J_0^2/\Delta^2)\,T$ is visible 
only for $|J_0| < 2 \Delta $. For higher interaction strengths $J_0$, $\gamma$ remains 
practically zero in the low-temperature regime. The maximal rate obtained  as $T \to \infty$ 
is given by \eqref{ratehT} : $\gamma_\infty = \sqrt{\pi} \lambda^2 [1+\exp(-J_0^2/\Delta^2)]/2\Delta $. 
For $|J_0| > 2 \Delta$, the high-temperature rate $\gamma_\infty$ is essentially independent 
of $J_0$ and the temperature dependence of $\gamma$ is well described by the weak disorder 
approximation $\gamma \simeq (\sqrt{\pi} \lambda^2 /2\Delta ) [ 1 - \tanh(\beta J_0)^2 ]$. Note that the agreement with 
this approximation  improves with increasing $J_0$.

We now consider a  uniform distribution for the intra-bath interaction strength: 
\begin{eqnarray}
p(J) &=& (2\sqrt{3}\Delta )^{-1} \quad  \mathrm{for} \quad  \vert J-J_0\vert < \sqrt{3}\Delta \nonumber \\
&=& 0 \quad ~~~~~~~~~~~~\mathrm{otherwise}  \quad .
\end{eqnarray} 
In this case, the integral \eqref{rate} can be evaluated exactly for any temperature $T$ :
\begin{widetext}\begin{eqnarray}
\gamma &=& \frac{\pi \lambda^2 T}{12\Delta^2} \left\{ \tanh \left[ \frac{|J_0|+\sqrt{3}\Delta}{T} \right] -  
3 \tanh \left[\frac{ |J_0|-\sqrt{3}\Delta}{T} \right]\right\} \quad \mathrm{for} \quad |J_0|< \sqrt{3} \Delta  \;\;\; \\
 &=& \frac{\pi \lambda^2 T}{12  \Delta^2} \left\{ \tanh \left[  \frac{|J_0|+\sqrt{3}\Delta}{T} \right] -  
\tanh \left[ \frac{|J_0|-\sqrt{3} \Delta}{T}\right]\right\} \quad \mathrm{for} \quad |J_0|>\sqrt{3} \Delta \nonumber \quad .
\end{eqnarray}  \end{widetext}
For $|J_0|<\sqrt{3} \Delta $, $\gamma \propto T$  at low temperatures with 
a slope $\pi \lambda^2/3\Delta^2$ independent of $J_0$  whereas, 
for $|J_0|>\sqrt{3} \Delta $ the low temperature behavior is not linear. 
For $|J_0| > 4 \Delta $, the agreement 
between the exact result and the weak disorder approximation 
$\gamma \simeq (\pi \lambda^2 /2 \sqrt{3} \Delta ) [ 1 - \tanh(\beta J_0)^2 ]$ is excellent. 
Moreover, 
in this regime, the Gaussian and uniform bond distributions cannot be distinguished 
($\sqrt{\pi/3} \simeq 1.02$). An interesting feature of our results is that a non-zero $J_0$ 
favours the coherence of the central spin via a robust short range ordering of the bath spins. 

Comparing the above results with those obtained   for a bath comprising free spins, 
we see that  the timescale of the decoherence generated by the spin chain bath  $\gamma_\infty^{-1} \sim \Delta /\lambda^2$  
(for weak coupling $\lambda \ll \Delta $)  is much longer 
 than the decoherence time  $\lambda^{-1}$ (for $\lambda \ll \sqrt{N}$)  obtained in the free case (see \aref{Isb}). 
 As we will demonstrate later, $\gamma_\infty^{-1} \gg \lambda^{-1}$ for any dimensionality 
of the spin bath lattice.      
 This clearly illustrates the fact that interactions 
in the bath significantly slow down  the decoherence of the central spin. 

Another interesting comparison is to an Ohmic boson bath.
Contrary to the Ising bath, the decoherence rate 
$\gamma_{bos}$ of an Ohmic boson bath is proportional to $T$ in the whole temperature range 
and thus does not saturate at high temperatures. This forces the question as to whether 
interactions between the
bosons also lead to a saturation of the rate $\gamma_{bos}$. This warrants further work which is
beyond the scope of the present paper.  If the central spin is coupled both to a spin 
bath and a boson bath, the resulting coherence $M(t)$ is given by the product of  \eqref{Mt} and 
a similar factor, with $H_B$ replaced by  the boson bath Hamiltonian $H_b$ and 
the interaction operator $V$ by an analogous boson operator $V_b$. Then in 
the Markovian regime  at weak coupling, the total decoherence rate is simply the sum $\gamma_{bos}+\gamma$. 
Consequently, at high temperatures the Ohmic boson bath dominates but for temperature 
$T \lsim \Delta$ the Ising bath has to be taken into account.  
The question of their relative dominance depends on the various bath coupling constants and
can vary from system to system.  

\subsubsection{Non-Markovian regime}

\begin{figure}
\centering \includegraphics[width=0.45\textwidth]{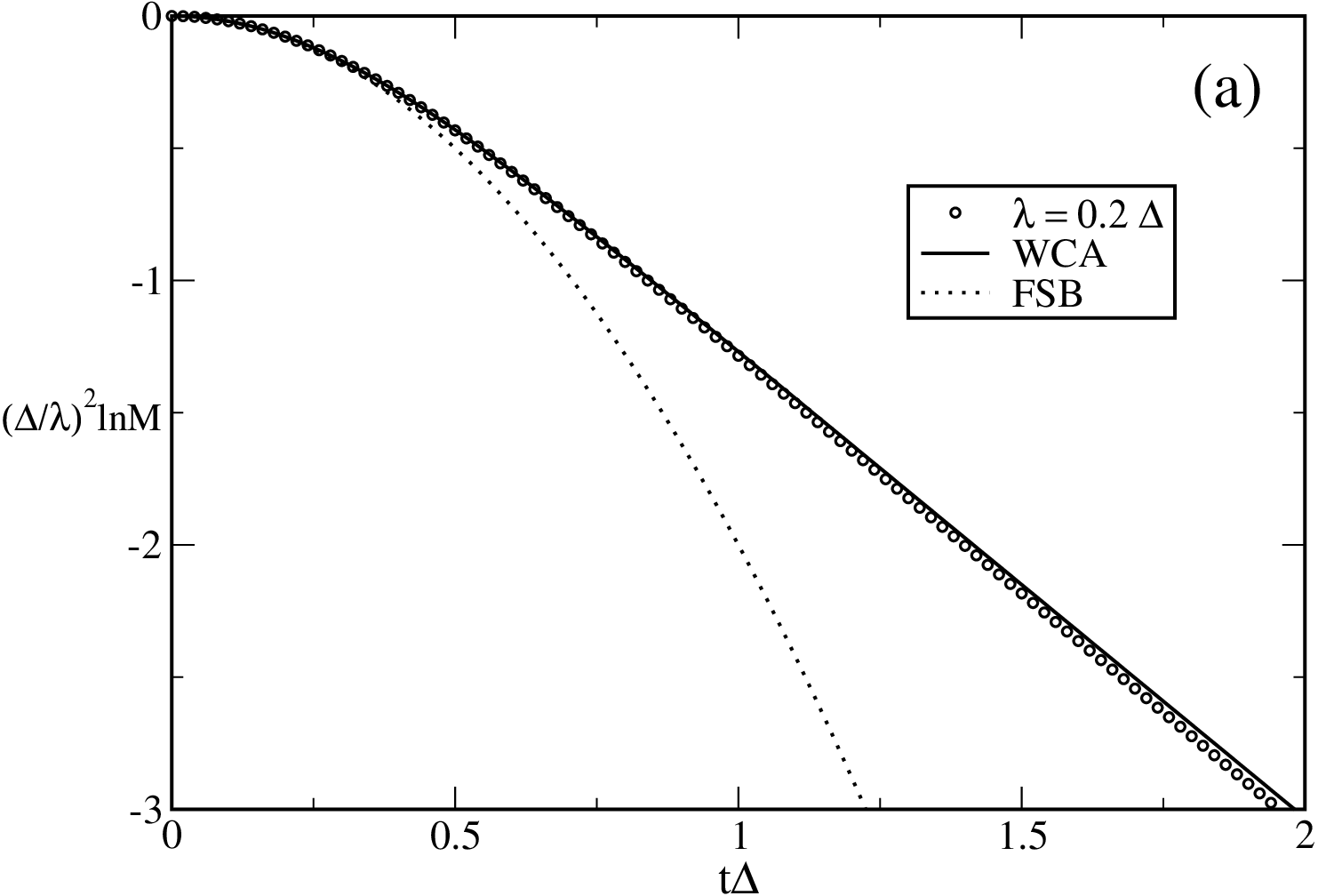}
\vskip0.9cm
\includegraphics[width=0.45\textwidth]{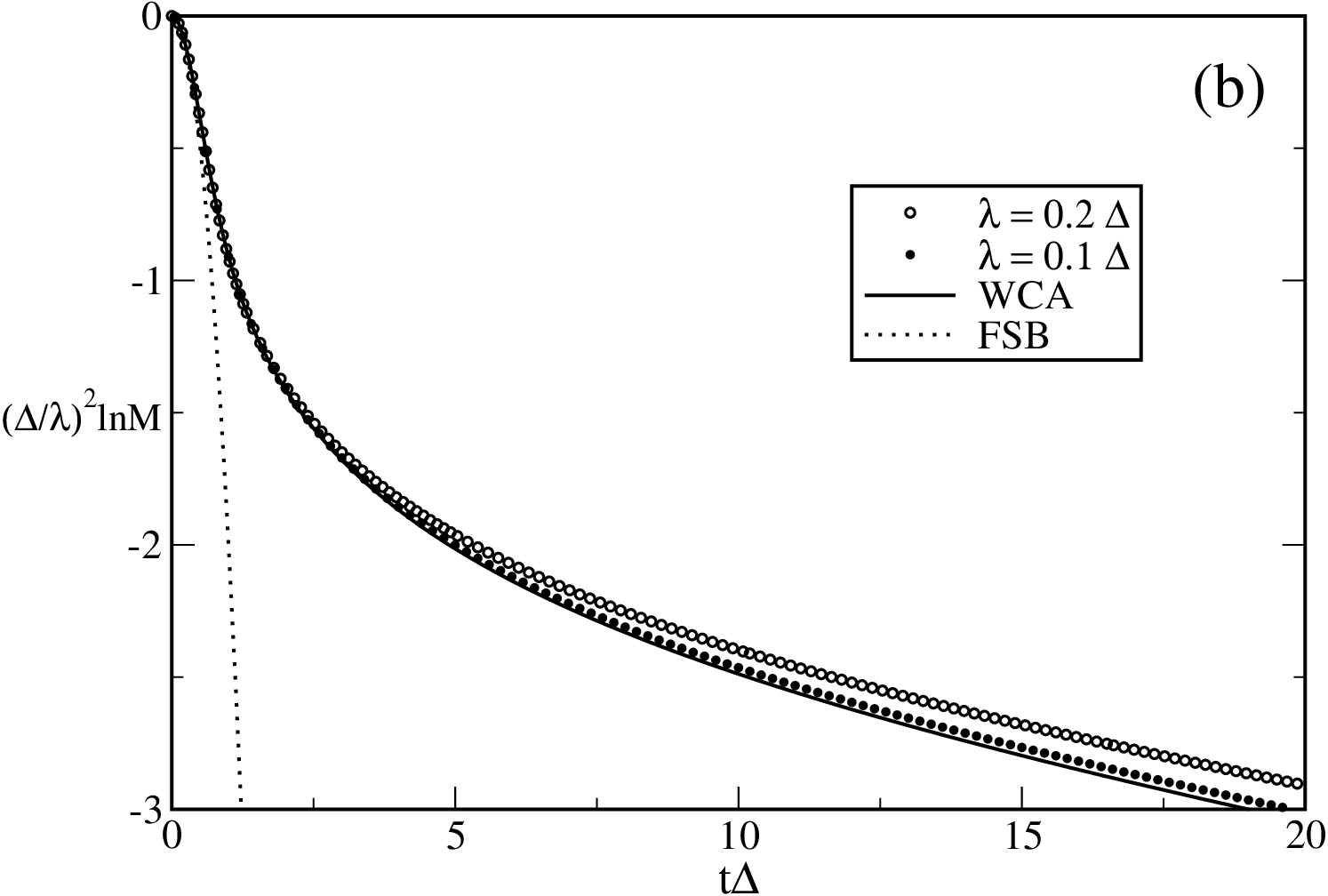}

\caption{$(\Delta/\lambda)^2 \ln M$ as a function of time $t\Delta$ 
within the Born approximation for a Gaussian bond distribution with zero
mean for coupling strengths  $\lambda=0.1 \Delta$, $0.2 \Delta$ 
and temperatures (a) $T =10 \Delta $ and (b) $T=0.02 \Delta$. 
The full line is the weak coupling approximation and the dotted line corresponds 
to a free spin bath.}
\label{fig:nonMarkov}

\end{figure}

Here, we discuss the non-Markovian aspects of the decoherence of the central spin 
essentially seen at low temperatures and at short and intermediate time scales. 
We first consider the case of a bond distribution with zero mean. 
The time evolution of the coherence $M(t)$ within the Born approximation, 
shown in  \fref{fig:nonMarkov}, is obtained by a numerical evaluation of $\Gamma_2(E)$, 
$\Lambda_2(E)$ and $M(t)$ using the expressions \eqref{Gamma1D}, \eqref{Ht} and \eqref{ThetaM} for a Gaussian
bond distribution \eqref{Gbd} with $J_0=0$.  
Note that the agreement with the weak coupling approximation \eqref{Mapprox} is remarkably good even for 
reasonably large values of $\lambda$ i.e., of the order of $0.1 \Delta$. 
Though, in  \fref{fig:nonMarkov}, we illustrated the equivalence of the Born and weak coupling approximations 
for $M(t)$ in the weak coupling regime, we nonetheless expect, based on the analyticity arguments presented 
in Sec. \ref{sec:wcr}, the weak coupling approximation \eqref{Mapprox} to be an exact description of the full 
coherence $M$  for low enough $\lambda$. 

We now discuss the influence of
 temperature on the behavior of the coherence $M(t)$. 
In  \fref{fig:nonMarkov}, we see that at high temperatures, $M$ remains
practically constant for times $t \lsim \Delta^{-1}$
and then decays exponentially
as described by \eqref{Mapproxlt}. For temperatures $T \lsim \Delta $,
the Markovian regime is preceded by a novel intermediate time regime $\Delta^{-1} \lsim t \lsim \beta$. 
The difference between the high-temperature
and the low-temperature decoherence can be traced back to the temperature dependence of
the function $\Gamma_2$. At high enough temperatures, 
since $\Gamma_2(E)$ is essentially a peak of width $\Delta$, one crosses over from
the short time regime to the Markovian regime for $t \simeq \Delta^{-1}$. 
As the temperature is lowered, $\Gamma_2(0)$ steadily decreases but the curvature $\Gamma''_2 (0)$ 
remains negative. However, below a certain
temperature $\Gamma''_2 (0)$ becomes positive, see  \fref{fig:gamma} and
the function $\Gamma_2$ can be effectively characterized by two energy scales: 
$T$ and $\Delta $. At low temperatures,
$\Gamma_2$ remains practically constant for $|E| \lsim T$, increases
linearly for larger energies with a slope $(\lambda/\Delta )^2$
and finally vanishes for $|E| \gsim \Delta $.
These three energy regimes result in three different time regimes for the coherence
\eqref{Mapprox}. The short-time behaviour ($t \lsim \Delta^{-1}$) is determined 
by the long-energy tails of $\Gamma_2$. For intermediate times  $\Delta^{-1} \lsim t \lsim \beta$, 
the linear regime of $\Gamma_2$ yields a power law decay of the coherence
$M(t) \propto t^{-\epsilon}$ where the exponent is given by
\begin{equation}
\epsilon=\frac{2}{\pi} \left( \frac{\lambda}{\Delta } \right)^2 \quad .
\end{equation}

For long times ($t \gsim \beta$) the integral \eqref{Mapprox} is dominated by the energies
$|E| \lsim T$ for which $\Gamma_2(E) \simeq \gamma$ and hence
the decoherence is essentially exponential with the rate $\gamma$.
However, one should be cautious about extending the above results to ultra-low temperatures 
because the contribution of the energies $|E| \gsim T$, given by 
$\Lambda_2^\prime(0)$, diverges in the limit $T \to 0$. 
This divergence is logarithmic with a prefactor $\epsilon$.
At zero temperature, the Markovian regime disappears and the coherence vanishes
in the limit $t \to \infty$ according to the power law $t^{-\epsilon}$.
The low temperature behavior of $M(t)$ obtained here is very similar to the
decoherence
induced by a boson bath in the strict Ohmic case \cite{QDS} i.e. for an Ohmic
spectral
function with a cut-off frequency $\omega_c$ far larger than $T$ and
$1/t$.
In this case, the coherence of the spin coupled to the boson bath is given by
$\ln M(t)=-K \ln[(\omega_c / T \pi) \sinh (\pi T t)]$ where $K$ is the
spin-bath
coupling strength. We recover a Markovian behavior, 
$\ln M(t) \simeq K \ln(2\pi T / \omega_c )-K\pi T t$, for times $t \gg
\beta$ and a power law, $M(t)=(\omega_c t)^{-K}$, at zero temperature.

We now present a more detailed comparison of our numerical results with the weak coupling 
approximation \eqref{Mapprox}, shown in  \fref{fig:nonMarkov}. 
For given $\lambda$ and $T$, the Born approximation deviates from the expression \eqref{Mapprox} 
as time increases. This difference, which is more pronounced at low temperatures, 
can be easily quantified in terms of higher order corrections 
stemming from the expansion of the analytically continued self-energy discussed in Sec. \ref{sec:NME}. 
At long times, the first correction to $\ln M(t)$ arises at $O(\lambda^4)$ and takes the form 
\begin{eqnarray}\label{correction}
\ln M(t) - \left[ \Lambda_2^\prime(0) - \Gamma_2(0) t \right] 
&\simeq&  - \Gamma_2(0)\Gamma_2^{\prime\prime}(0) + 
{\textstyle \frac{1}{2}} [\Lambda_2^\prime(0)]^2  \nonumber \\
&&
- \Gamma_2(0) \Lambda_2^\prime(0) \, t  \quad . 
\end{eqnarray}
In the high temperature regime, since $\Lambda_2^\prime(0)$ is positive, the weak coupling approximation 
given by the second-order terms in \eqref{correction} 
slightly overestimates $\ln M$. As the temperature is lowered, the evolution of the function $\Gamma_2$ discussed earlier (see  \fref{fig:gamma}) results in a sign change of $\Lambda_2^\prime(0)$. Thus for low enough 
temperatures, $\Lambda_2^\prime(0)$ is negative and the weak coupling approximation underestimates $\ln M$. 
This analysis is indeed very consistent with our results shown in  \fref{fig:nonMarkov}. 
This discussion clearly highlights the efficiency of our approach based on the analyticity properties 
of the self-energy, obtained within any approximation, to describe the long-time decoherence. 
We remark that other fourth order corrections to $\ln M(t)$ exist beyond the Born approximation.

We now consider the case $J_0 \neq 0$. For weak disorder 
$|J_0| \gg \Delta$, the function $\Gamma_2$ at low temperature consists 
of two Gaussian peaks of width $2\sqrt{2}\Delta$ centered around $\pm 4J_0$.
The low-temperature self-energy in this regime is thus very different
from that for $J_0=0$, see  \fref{fig:gamma}. Nonetheless, the qualitative 
behavior of the coherence $M$ at intermediate and long times, determined 
by the low-energy $\Gamma_2$, is very similar 
to the one discussed above for $J_0=0$.  For short times $t \lsim \Delta^{-1}$, 
contrary to the $\ln M(t) \propto -t^2$ behavior seen for  $J_0=0$, here 
the coherence $M(t)$ oscillates at the frequency $2J_0/\pi$. 
At zero temperature,
the long-time decoherence decays as a power law,  $M(t) \propto t^{-\eta}$ where the exponent 
$\eta=\epsilon \exp(-J_0^2/\Delta^2)$. As seen earlier for the Markovian decay 
rate $\gamma$, the low-temperature behavior seen here also signals a slowing down 
of the decoherence by a nonzero $J_0$.

\subsection{Higher-dimensional spin lattices}

Here we consider other geometries for an Ising spin bath described by
the Hamiltonian $H_B=-\sum_{\langle i j \rangle} J_{ij} \sigma_i^z \sigma_j^z$ where the
spins
occupy the sites $i$ of a regular lattice of arbitrary dimensionality and are coupled
by nearest-neighbor interactions. The bonds $J_{ij}$ 
are drawn independently from the Gaussian distribution $p$  with mean $J_0$ and variance
$\Delta^2$ given by \eqref{Gbd}. An important difference between the spin chain model and 
the generic Ising model on higher dimensional lattices is the presence of frustration 
arising from geometric constraints and/or randomness. 
Though frustration can give rise to novel ground states and related dynamical behaviour,
it renders any analytic study of these models very difficult. 
In this section, we focus on the high temperature regime where one can use a controlled
analytical method like the high temperature series expansion 
to study the effect of higher dimensions on the  decoherence.

To obtain the coherence $M$ at high temperatures, we expand the second-order self-energy 
\eqref{soSigma} in terms of the inverse temperature $\beta$. To do so, we first rewrite the time-dependent 
correlation \eqref{VV1} as 
\begin{widetext}
\begin{equation}
\mathrm{Re}\langle V(t)V\rangle = \frac{1}{Z'} \sum_k \lambda_k^2   \sum_{\{\sigma_i\}}\prod_{\langle i j \rangle } (1+\sigma_i \sigma_j \kappa_{ij})\mathrm{Re} 
\prod\nolimits_i^{(k)} \big[ \cos(2tJ_{ik})+i\sigma_i \sigma_k \sin(2tJ_{ik}) \big] \label{VtVhT}
\end{equation}\end{widetext}
where $\prod\nolimits_i^{(k)}$ denotes a product over the nearest neighbors of site $k$, 
$Z'=\sum_{\{\sigma_i\}} \prod_{\langle ij \rangle} (1+\sigma_i \sigma_j \kappa_{ij})$ and $\kappa_{ij}=\tanh(\beta J_{ij})$. 
Since 
$\kappa_{ij} \to 0$ as $\beta \to 0$,  we consider the above expression  as a power series 
in $\kappa_{ij}$. Multiplying out the two products in Eq.\eqref{VtVhT} generates a series 
of products of nearest-neigbour spin pairs\,: $(\sigma_i \sigma_j \sigma_k \sigma_l \ldots)$. 
Since Eq.\eqref{VtVhT} 
involves sums over configurations $\{\sigma_i\}$, each of these spin pair products contributes 
only if it simplifies to $1$. This implies, for instance, that the $n^\mathrm{th}$ order 
terms in the expansion of $Z'$ in $\kappa_{ij}$ correspond to closed loops comprising 
$n$ bonds on the lattice. An immediate consequence is that $Z'=1$ up to third order. 
However, in the expression \eqref{VtVhT} there exists another type of relevant spin pair products 
which involve repeated bonds. We discuss these terms in the following.

As $T \to \infty$, for equal couplings to the central spin $\lambda_k=\lambda N^{-1/2}$, 
the time-dependent correlation \eqref{VtVhT} becomes 
$\mathrm{Re}\langle V(t)V\rangle = \lambda^2 N^{-1} \sum_k \prod\nolimits_i^{(k)} \cos(2tJ_{ik}) $. 
Note that this correlation remains unchanged under $J_{ik} \to -J_{ik}$. This invariance 
stems from the equiprobability of every spin configuration at infinite temperature. This infinite-temperature correlation is self-averaging in the thermodynamic limit 
and hence the corresponding second order self-energy is given by the average of \eqref{soSigma} 
over the bond distribution. This leads to
\begin{equation}
\Gamma_2(E) =  2 \lambda^2 \int_{-\infty}^{\infty} \!\! dt \; e^{iEt} e^{-2\Delta^2 t^2 s} \cos(2J_0
t)^s
\label{GammahT}
\end{equation}
where $s$ is the coordination number of the lattice. As remarked earlier for the 1D case,
here also $\Gamma_2$ is the same for $\pm J_0$ for bond distributions symmetric around 
their mean values. Eq.\eqref{GammahT} shows that in the infinite temperature limit 
the only characteristic of the bath lattice which intervenes in the decoherence is $s$. 
In particular, there is no explicit dependence on the dimensionality. However, we shall 
show later that the detailed geometric characteristics of the lattice manifest themselves in 
the higher order corrections. Though static properties 
of the spin system are independent of $J_0$ and $\Delta$ in the infinite temperature limit, 
these parameters strongly influence 
the correlation $\mathrm{Re}\langle V(t)V\rangle$ and hence the decoherence. 
Using  \eqref{GammahT}, for the case $J_0=0$, we obtain from \eqref{Mapprox}
\begin{equation}
\ln M(t) =  - 2 \sqrt{\pi} \lambda^2 \tau t \, \mathrm{erf} (t/\tau)
+ 2 \lambda^2 \tau^2 \left( 1-e^{-t^2/\tau^2} \right)
\end{equation}
where $\mathrm{erf}$ is the error function and the characteristic time $\tau$ is defined by
\begin{equation}
\tau= \Delta^{-1} (2s)^{-1/2} \quad .
\end{equation}
For times $t \gg \tau$, we recover the Markovian regime where 
$\ln M(t) \simeq - 2 \sqrt{\pi} \lambda^2 \tau t + 2 \lambda^2 \tau^2$.
For $t \ll \tau$, one finds the usual short time
evolution\,:
$\ln M(t) \simeq - 2 \lambda^2 t^2$. The same time regimes
exist for $J_0 \neq 0$. For short times $t \ll \tau$ we find, for an even $s$,
\begin{equation}
\ln M(t) =  - \frac{\lambda^2}{2^{s-2} J_0^2} \sum_{n=0}^{s/2-1}
\frac{s! \sin[J_0 t(2n-s)]^2 }
{n!(s-n)! (2n-s)^2} -
\frac{s! \lambda^2 t^2}{2^{s-1}[(s/2)!]^2} \quad . \label{sthT}
\end{equation}
If $s$ is an odd number, the short-time decoherence is described by
\eqref{sthT} without the quadratic term in $t$ and with a sum running from $0$ to $(s-1)/2$. 
Clearly the effect of a nonzero $J_0$ is to induce oscillations in the coherence $M(t)$. 
The decoherence rate in the Markovian regime is given by
\begin{equation}
\gamma_\infty =  \frac{\lambda^2}{2^{s-1} \Delta }  \sqrt{\frac{\pi}{2s}}
\sum_{n=0}^{s} \frac{s!}{n!(s-n)!} e^{-J_0^2(2n-s)^2/2s \Delta^2} \quad .
\label{ratehTs}
\end{equation}
For $s=2$ this expression simplifies to the result \eqref{ratehT} obtained for the spin chain. 
As anticipated, the decoherence time $\gamma_\infty^{-1}$ is of the order of $\Delta/\lambda^2$ 
and hence far longer than that for free spins which is of the order of $\lambda^{-1}$. 
The infinite-temperature rate manifests the influence of the lattice geometry:
for lattices with an odd $s$ like the honeycomb lattice ($s=3$), $\gamma_\infty \to 0$  in 
the weak disorder limit $\Delta  \ll |J_0|$ whereas  for an even $s$, triangular and square 
lattices for example, $\gamma_\infty \neq 0$  in this limit.

We now evaluate the leading order corrections in $\beta$ to the decoherence rate $\gamma$.  
To illustrate the significance of the lattice geometry we consider various bidimensional lattices. 
For a triangular lattice, the lowest order terms in the sum over the configurations $\{\sigma_i\}$ 
in \eqref{VtVhT} take the form $-\sin(2tJ_{ik})\kappa_{ij}\sin(2tJ_{jk})$ where
the site $i$ is a neighbor of the site $k$ and the site $j$ is a neighbor of
both sites $k$ and $i$. Consequently, the rate up to first order reads
\begin{widetext}\begin{equation}
\gamma^T=\gamma^T_\infty - {\frac{\sqrt{3\pi}}{16}} \frac{\lambda^2 \beta
J_0}{\Delta }  \left[ 2-2e^{-4J_0^2/3\Delta^2}+e^{-J_0^2/3\Delta^2}
-e^{-3J_0^2/\Delta^2} \right]+O(\beta^2)
\end{equation}\end{widetext}
where $\gamma^T_\infty$ is given by \eqref{ratehTs}
with $s=6$. The explicit appearence of $J_0$ in this expansion is 
linked to the fact that the first nonzero correction for this lattice 
occurs at first order in $\beta$. Moreover, since this correction is positive (negative) for baths 
with a majority of antiferromagnetic (ferromagnetic) bonds, the ensuing decoherence time 
is longer for a predominantly ferromagnetic bath.  This clearly highlights the importance of frustration 
in determining decoherence. Thus at high temperatures, the decoherence induced by the triangular lattice 
is very different than that by the linear chain\,: the convergence of $\gamma^T$ to $\gamma^T_\infty$ 
is slower and depends on the sign of $J_0$. 
We also remark that though the cubic and triangular lattices have the same coordination $s=6$ 
and hence the same infinite-temperature rate, the corrections are different since there is no 
three-bond closed loop on the cubic lattice. 
For the honeycomb lattice, the first corrections arise at second order. These corrections 
correspond to loops with two repeated bonds\,: $-\kappa_{ki}\sin(2tJ_{ik})\kappa_{kj}\sin(2tJ_{jk})$. 
Here the contribution of closed loops becomes relevant at higher orders. The resulting 
decoherence rate is 

\begin{widetext}\begin{equation}
\gamma^H=\gamma^H_\infty - \sqrt{\frac{\pi}{24}}
\frac{\lambda^2 \beta^2 }{\Delta } \left[
\left( 3{\Delta^2}  + 4 J_0^2 \right) e^{-J_0^2/6\Delta^2} 
+{\Delta^2} e^{-3J_0^2/2\Delta^2}  \right]+O(\beta^4)
\end{equation}\end{widetext}
where $\gamma^H_\infty$ is given by \eqref{ratehTs} with $s=3$.
Finally for the square lattice we obtain the decoherence rate
\begin{widetext}\begin{equation}
\gamma^S=\gamma^S_\infty - \frac{1}{4} \sqrt{\frac{\pi}{2}} \frac{\lambda^2 \beta^2
}{\Delta}
\left[\left( \frac{3}{4} \Delta^2-2 J_0^2 \right) e^{-2J_0^2/\Delta^2}
+ 3(\Delta^2+J_0^2) e^{-J_0^2/2\Delta^2}+
\frac{9}{4}\Delta^2+5 J_0^2 \right]+O(\beta^4) 
\end{equation}\end{widetext}
where $\gamma^S_\infty$ is given by \eqref{ratehTs} with $s=4$. 
We mention that for the square lattice both closed loops and loops with repeated bonds 
contribute to the lowest order correction. Interestingly, the convergence 
of the rates $\gamma^H$ and $\gamma^S$ to their respective infinite-temperature limits 
is reminiscent of that of the chain. Moreover, as seen for the spin chain, 
our results for $\gamma^H$ and $\gamma^S$ are independent of the sign of $J_0$. 
This feature can be attributed to the bipartite nature of these lattices. 
As explained for the chain, the respective local field operators and Hamiltonians 
are invariant under a sign change of the bonds $J_{ij}$ coupled with 
a suitable unitary transformation for the spins. Since this argument is valid in the entire 
paramagnetic phase, the higher order corrections to $\gamma^H$ and $\gamma^S$ 
are expected to be independent of the sign of $J_0$ in this phase. On the other hand, due to its 
non-bipartite nature no analoguous transformation exists for the triangular lattice.
It would be interesting to study the evolution of  these rates behave as one lowers temperature and enters
a non--paramagnetic phase.

\subsection{Infinite-ranged Ising bath}

We now consider a bath described by the  mean field Sherrington-Kirkpatrick model 
where all the Ising  spins interact with each other. We restrict ourselves to the well studied 
case of spin-spin interaction strengths distributed around a zero mean with a variance $\Delta^2$. 
This model exhibits a spin glass phase transition at a
finite temperature $T_{sg}=\Delta$ below which the spins freeze \cite{sk}.  The spin glass phase is
characterized by the large number of metastable states present which then lead to
anomalous dynamic behaviour. 
 An interesting question is whether the
decoherence manifests novel features as goes from the high temperature paramagnetic
phase to the low temperature spin glass phase.  

 As shown in \sref{sec:DIsb}, the coherence $M(t)$ of the central spin in the weak coupling regime 
can be obtained through a knowledge of the local-field distribution $P(h)$. 
Though the infinite-range model  has been extensively  studied in the past using the replica approach 
and numerical simulations, it is not easy to obtain the local-field distribution for all fields and temperatures. 
Here, we do not delve into the problem of calculating $P(h)$ but use existing results \cite{salfd} 
to make predictions for the coherence  of the central spin coupled to this mean field bath.
In the high temperature paramagnetic phase, the local-field distribution is 
\begin{equation}
P(h)= \frac 1{2\Delta \sqrt{2\pi}} \left[e^{-(h-\beta \Delta^2)^2/2\Delta^2}
+e^{-(h+\beta \Delta^2)^2/2\Delta^2} \label{phhT} \right] \quad .
\end{equation}
In the spin glass phase, calculations based on the replica formalism
yield  the following result for $P(0)$  provided the temperature
$T \simeq T_{sg} $:
\begin{equation}
P(0)=  \frac 1{\Delta \sqrt{2\pi e}} \left[ 1- \left(1- \frac{T}{T_{sg}} \right) -\frac 1 3 \left( 1- \frac{T}{T_{sg}} \right)^4  \right] + ... \quad .
\end{equation}
These analytic results are sufficient to determine the decoherence in the Markovian regime\,: 
the decoherence rate is given by 
$\gamma = \sqrt{2\pi} \lambda^2 \exp[-\frac 1 2 (T_{sg}/T)^2]/\Delta$ for $T > T_{sg}$ and 
$\gamma = \sqrt{2\pi/e} \lambda^2  [ 1- (1- T/T_{sg}) -\frac 1 3 (1- T/T_{sg})^4 ]/\Delta$ for temperatures in 
the vicinity of the spin glass transition temperature i.e., $ T \simeq T_{sg}$ . 
Clearly, in the Markovian regime, one does not see any sign of the spin glass transition. 
Note that $\gamma$ saturates at infinite temperature to a value comparable to that obtained earlier 
for finite-ranged lattice models. This implies  that even in the case of a  maximally frustrated bath, 
the central spin decoheres at timescales longer than those for free spins.  

For lower temperatures, only numerical solutions exist for the local-field distribution \cite{salfd,Plefka}. 
These results suggest a continuous variation of $P(h)$ with temperature. The only significant signature 
of the transition is a flattening of $P$ at $T=T_{sg}$ which has no manifest effect on the decoherence. 
Moreover, numerical extrapolations of $P(0)$ to low temperatures indicate a rate $\gamma \sim (\lambda/\Delta)^2 \, T$ 
which is very similar to the linear $T$ behaviour seen in the Ising spin chain system at low enough
temperatures.  Again,  since $\gamma=0$ at zero temperature, the decoherence is no longer Markovian. 
The form of $M(t)$ is then dictated by the low-energy behavior of $\Gamma_2$ through \eqref{Mapprox}. 
Based on the numerical inference $P(h) \propto h$ for $T=0$, the central spin is expected to decohere 
as a power law at $T=0$ \cite{salfd,Plefka}. To conclude, we see that both the thermal transition and 
the spin glass order of the Sherrington-Kirkpatrick model do not have any palpable effect 
on the asymptotic decoherence in the weak coupling regime.       

\section{Conclusions}

In this paper, we have studied the decoherence induced by a Ising spin bath with 
random intra-bath interactions. The resolvent operator formalism was used to determine 
the coherence $M$ of the central spin for weak coupling to the bath. 
We  then obtained detailed analytical results for the disordered Ising spin chain bath for arbitrary 
temperature.  
The decoherence was found to be independent of the sign of the mean 
intra-bath interaction strength $J_0$ for symmetric bond distributions. Three regimes were  identified 
in the time evolution of $M(t)$\,: a short time Gaussian decay, an intermediate time power law behavior 
and the usual asymptotic Markovian regime. The relative sizes of these regimes are fixed by temperature. 
At zero temperature, the Markovian regime was found to vanish  and the decoherence is essentially described by 
a power law decay. We also studied the decoherence induced by an infinite-ranged Ising spin glass bath 
and Ising models on lattices in dimensions greater than one. For all these baths, the Markovian rate 
was found to saturate to a finite value at infinite temperature, which is much smaller than the corresponding rate 
for a free spin bath. Our  results clearly indicate  that  intra-bath interactions significantly increase the timescales 
over which the central spin decoheres.  

For the infinite-ranged Ising spin glass, our conclusions  based on existing results 
suggest that the thermal spin glass transition has no visible effect on the decoherence. Plausibly this is an artefact of 
the infinite-ranged interactions and/or the Ising nature of the spins. This raises the general issue of 
the influence of thermal and quantum phase transitions and the resulting orders in finite-ranged spin baths 
on the decoherence of the central spin. In most realistic cases, the spin environment consists of 
Heisenberg spins. In this case, one expects the dynamics of the bath to be richer and this may 
have interesting consequences for the decoherence. This however, is beyond the scope of the analytic 
work presented in this paper. 
An interesting question is the effect of a strong coupling 
between the central spin and the bath. For a bath of independent spins/bosons the results obtained 
in the weak coupling regime 
are qualitatively valid even for strong coupling. However, in the presence of interactions, this is not necessarily 
the case and one can expect novel dynamical behaviors. 
A natural extension of our work would be to include the intrinsic dynamics
 of the central spin and study  the  relaxation
induced by the spin bath.
These and other questions are left for future work. 

\begin{acknowledgments}
R.C. acknowledges support from ACI Grant No. JC-2026.
\end{acknowledgments}

\begin{appendix}

\section{Independent spin bath}\label{Isb}

Here we derive an exact expression for the coherence $M(t)$ given by \eqref{Mt} 
in the case of a bath comprising non-interacting spins described by the Hamiltonian $H_B=-\sum_i h_i \sigma_i^z$.
Since the spins are independent, the trace \eqref{Mt} can be factorized as 
\begin{equation}
M(t)=\prod_{k=1}^N \frac{ \mathrm{Tr} \left[ 
e^{\beta h_k \sigma_k^z}e^{-it( h_k \sigma_k^z - \lambda_k \sigma_k^x )}
e^{it( h_k \sigma_k^z + \lambda_k \sigma_k^x )}   \right]}{2\cosh (\beta h_k)} \quad . 
\end{equation} 
To evaluate the factors in this expression we require the diagonal elements 
of $2\times 2$ matrices of the form $\exp[-ibA(a)]\exp[ibA(a)]$ where 
$A(a)=a\sigma^z+\sigma^x$. Diagonalizing $A(a)$, we find that these diagonal elements 
are equal and given by
\begin{equation}
\langle \sigma | e^{-ibA(a)} e^{ibA(a)} | \sigma \rangle  = 
\frac{1}{1+a^2} \left[ a^2+ \cos\left( 2b \sqrt{1+a^2}\right) \right] \quad .
\end{equation} 
The resulting  coherence $M(t)$ is independent of the temperature and 
can be written  as
\begin{equation}
\ln M(t) = \sum_{k=1}^N \ln \left[ 
1- \frac{2\lambda_k^2}{\lambda_k^2+h_k^2} \sin \left( t \sqrt{\lambda_k^2+h_k^2}\right)^2 \right] \quad .
\end{equation}
\noindent
For  $h_k=0$, the  above expression leads to the coherence, $M(t)
=\prod_k \cos (2 \lambda_k t)$ which then culminates in a  Gaussian decay  $M(t)=\exp(-2t^2 \sum_k \lambda_k^2)$ 
for weak coupling to the bath.
For nonzero local fields $h_k$ the coherence in the   weak coupling  regime reads:
 \begin{equation}
\ln M(t) = -2 \lambda^2 \int dh \; P(h) \frac{\sin(ht)^2}{h^2}
\end{equation}
where the coupling strength $\lambda$ and  the field distribution $P$ are defined by
$\lambda^2= \sum_k \lambda_k^2$ and  $P(h)=\sum_k (\lambda_k/\lambda)^2 \delta(h-h_k)$. 
This result is the same as that obtained earlier in \eqref{Mapprox} for the specific 
bath considered here. 

\section{Derivation of the self-energy}\label{Dse}

In this Appendix,  we derive the expressions \eqref{LT} and \eqref{Sigma} for the Laplace transform ${\tilde M}(z)$ 
of the coherence \eqref{Mt} and we show that ${\tilde M}(z)$ is analytic in the upper and lower half 
planes. 
We first rewrite the expression \eqref{Mt} as 
\begin{equation}
M(t)=\mathrm{Tr} \left( e^{-i{\cal L} t} \rho_B \right)
\end{equation}
where ${\cal L}$ is a superoperator in the bath Liouville space defined by 
${\cal L} A = (H_B+V) A - A (H_B-V)$ where $A$ is any operator in the bath Hilbert space. 
The Laplace transform \eqref{LTdef} can then be written as ${\tilde M}(z)=\mathrm{Tr} [G(z) \rho_B]$ 
where $G(z)=(z-{\cal L})^{-1}$ is the resolvent of the operator ${\cal L}$. 

We now introduce the superoperators ${\cal P}$ and ${\cal Q}$ 
defined by ${\cal Q}=1-{\cal P}$ and ${\cal P} A=\mathrm{Tr} (A) \rho_B $ where $A$ is any 
operator in the bath Hilbert space. Since $\mathrm{Tr} ( \rho_B) =1$, ${\cal P}$ and ${\cal Q}$ 
are projection operators. 
Using ${\cal P}$, ${\cal Q}$ and $(z-{\cal L})G(z)=1$, we obtain the following coupled equations 
\begin{eqnarray} 
{\cal P}(z-{\cal L}){\cal P}\,{\cal P}G(z) \rho_B + {\cal P}(z-{\cal L}){\cal Q} \,{\cal Q}G(z) \rho_B\!\!&=&\!\!\rho_B \nonumber \\
{\cal Q}(z-{\cal L}){\cal P}\,{\cal P}G(z)\rho_B + {\cal Q}(z-{\cal L}){\cal Q}\,{\cal Q}G(z)\rho_B\!\!&=&\!\!0
\end{eqnarray}
for the operators ${\cal P}G(z) \rho_B$ and ${\cal Q}G(z) \rho_B$. Solving the latter
for ${\cal Q}G(z) \rho_B$ in terms of ${\cal P}G(z) \rho_B$ and then substituting in the former  yields 
\begin{equation}
{\cal P} \left[ z-{\cal L} - {\cal L} {\cal Q} (z-{\cal Q}{\cal L}{\cal Q})^{-1}
{\cal Q} {\cal L} \right] {\cal P}G(z) \rho_B =  \rho_B \quad .
\end{equation}
Finally tracing both sides gives $[z-\Sigma(z)]{\tilde M}(z)=1$ where $\Sigma(z)$ is given by \eqref{Sigma}.

We now discuss the analyticity properties of the Laplace transform ${\tilde M}(z)$. 
Consider an eigenoperator $A$ of ${\cal L}$ with the eigenvalue $\lambda$ : ${\cal L} A = \lambda A$. 
The scalar product $\mathrm{Tr}(A^{\dag} {\cal L} A)=\lambda \mathrm{Tr}(A^{\dag} A)$ can also be written 
\begin{eqnarray}
\mathrm{Tr}\left( A^{\dag}  {\cal L} A \right)&=&\mathrm{Tr}
\left[ A^{\dag}(H_B+V)A - (H_B-V)A^{\dag} A \right]^*\nonumber \\ 
&=& \mathrm{Tr}\left(A^{\dag} {\cal L} A \right)^*=\lambda^* \mathrm{Tr}(A^{\dag} A) 
\end{eqnarray}
where the first equality is obtained using the Hermiticity of $H_B$ and $V$ and the second one 
using the cyclic property of the trace. Consequently, the eigenvalues of ${\cal L}$ are real 
and thus the resolvent $G(z)$ and the Laplace transform ${\tilde M}(z)$ are analytic in the upper and lower half planes.

\section{Second-order self-energy}\label{Sose}

Here, we show that the second-order self-energy $\Sigma_2$  can 
be rewritten in terms of the time-dependent correlation of the interaction 
operator $V$ as given by \eqref{soSigma}. Let $| \alpha \rangle$ denote the eigenstates of the Hamiltonian $H_B$ and 
$|\alpha, \beta ) = | \alpha \rangle \langle \beta |$ denote the eigenstates of the corresponding Liouville operator 
${\cal L}_B$. Any superoperator 
${\cal F}$ in the Liouville space can be expanded in this eigenbasis as 
\begin{equation}
{\cal F}= \sum_{\alpha, \beta, \gamma, \delta}  ( \alpha, \beta | {\cal F}|\gamma, \delta ) \; |\alpha, \beta )( \gamma, \delta |
\end{equation} 
where the scalar product in the Liouville space is defined by $(A|B)=\mathrm{Tr} (A^{\dag} B)$. 
The following decompositions are useful for our purpose :
\begin{eqnarray}
&&{\cal P} = \rho_B \sum_{\beta} ( \beta , \beta | = \sum_{\alpha, \beta} \langle \alpha | \rho_B | \alpha \rangle  |\alpha, \alpha ) ( \beta , \beta | 
 \\
&&(z-{\cal Q}{\cal L}_B {\cal Q})^{-1} = \sum_{\alpha, \beta} \frac{1}{z-E_\alpha+E_\beta}  |\alpha, \beta ) ( \alpha , \beta | \nonumber \\
&&{\cal L}_V =  \sum_{\alpha, \beta,\gamma} \Big[ \langle \alpha | V | \gamma \rangle  |\alpha, \beta ) ( \gamma , \beta | + 
 \langle \gamma | V | \beta \rangle  |\alpha, \beta ) ( \alpha , \gamma | \Big] \nonumber
\end{eqnarray}  
where ${\cal P}=1-{\cal Q}$ and $E_\alpha$ is the eigenergy corresponding to the eigenstate 
$| \alpha \rangle$.
Using these results we find
\begin{eqnarray}
\Sigma_2 (z)  =&&  2 \sum_{\alpha, \beta}  \langle \alpha | \rho_B | \alpha \rangle |\langle \alpha
 | V | \beta \rangle|^2  \\
&&~~ \times  \left[ \frac{1}{z-E_\alpha+E_\beta}   
 + \frac{1}{z-E_\beta+E_\alpha}  \right]
\quad .\nonumber
\end{eqnarray} 
Comparing this expression with the time-dependent correlation 
\begin{eqnarray}
\langle V(t)V \rangle& =& \mathrm{Tr} \left( \rho_B e^{itH} V e^{-itH} V \right) \nonumber \\ 
&=& \sum_{\alpha, \beta}  \langle \alpha | \rho_B | \alpha \rangle |\langle \alpha | V | \beta \rangle|^2 e^{it(E_\alpha-E_\beta)}
\end{eqnarray} 
we infer that the Laplace transform of $4\mathrm{Re}\langle V(t)V \rangle$ is $\Sigma_2 (z)$ as given by \eqref{soSigma}. 
\end{appendix}


\begin{thebibliography}{99}



\bibitem{Caldeira-Leggett}
A. Caldeira and A. Leggett, Phys. Rev. Lett. {\bf 46}, 211 (1981); 
Ann. Phys. {\bf 149}, 374 (1983).

\bibitem{QDS} U. Weiss, {\it Quantum dissipative systems} (World Scientific, Singapore, 1993).

\bibitem{stamp-review}
N.V. Prokof'ev and P.C.E. Stamp, Rep. Prog. {\bf 63}, 669 (2000).

\bibitem{Coish-Loss}
W. Coish and D. Loss,Phys. Rev. B {\bf 70}, 195340 (2004). 

\bibitem{Marcus}J. M. Taylor, H.-A. Engel, W. Dur, A. Yacoby,
C. M. Marcus, P. Zoller, and M. D. Lukin, Nature Physics{\bf  1}, 177 (2005); J. M. Taylor, J. R. Petta, A. C. Johnson, A. Yacoby, C. M. Marcus and  M. D. Lukin, cond-mat/0602470 (2006).

\bibitem{Hu}
X. Hu, cond-mat/0411012 (2004).





\bibitem{Hanggi} J. Shao and P. H\"anggi, Phys. Rev. Lett. {\bf 81}, 5710 (1998). 

\bibitem{Hanggi1} K.M. Forsythe and N. Makri, Phys. Rev. B {\bf 60}, 972 (1999).

\bibitem{italian-meanfield}
S. Paganelli, F. de Pasquale and S.M. Giampaolo, Phys. Rev. A {\bf 66}, 052317 (2002).

\bibitem{sib} L. Tessieri and J. Wilkie, J. Phys. A : Math. General. {\bf 36}, 12305 (2003). 

\bibitem{Yuan} X.-Z. Yuan and K.-D. Zhu, Europhys. Lett., {\bf 69}, 868 (2005).

\bibitem{cb} J. Lages,
V. V. Dobrovitski, M. I. Katsnelson, H. A. De Raedt, and B. N. Harmon, Phys. Rev. E {\bf 72}, 026225 (2005).

\bibitem{spinglassbook}
 K.H. Fischer and J.A. Hertz,{\it Spin Glasses} (Cambridge University Press, 1991). 
  
\bibitem{LE} R.A. Jalabert and H.M. Pastawski, Phys. Rev. Lett. {\bf
86}, 2490 (2001).
\bibitem{LE1}
Z.P. Karkuszewski, C. Jarzynki and W.H. Zurek, Phys. Rev. Lett. {\bf 89}, 170405 (2002). 

\bibitem{fazio}D. Rossini, T. Calarco, V. Giovannetti, S. Montangero
and R. Fazio, cond-mat/065051.

\bibitem{ROM} S. Dattagupta, H. Grabert and R. Jung, J. Phys. :
Cond. Mat. {\bf 1}, 1405 (1989)

\bibitem{ROM1} C. Cohen-Tannoudji, J. Dupont-Roc and G. Grynberg, {\it Processus d'interaction entre photons et atomes} (CNRS Editions, Paris, 1988).


\bibitem{lfd} M. Thomsen, M.F. Thorpe, T.C. Choy and D. Sherrington, Phys. Rev. B {\bf 30}, 250 (1984). 

\bibitem{salfd} M. Thomsen, M.F. Thorpe, T.C. Choy, D. Sherrington and H.J. Sommers, 
Phys. Rev. B {\bf 33}, 1931 (1986). 

\bibitem{Plefka}
T. Plefka,  Phys. Rev. B {\bf 65} 224206 (2002).

\bibitem{1Dlfd} M. Barma, Solid State Commun. {\bf 30}, 11 (1979).


\bibitem{sk}
D. Sherrington and S. Kirkpatrick, Phys. Rev B. 
{\bf 17}, 4384 (1978).


\end{thebibliography}
\end{document}